\documentclass[pdflatex,sn-mathphys-num]{sn-jnl}

\usepackage{graphicx}%
\usepackage{multirow}%
\usepackage{amsmath,amssymb,amsfonts}%
\usepackage{amsthm}%
\usepackage{mathrsfs}%
\usepackage[title]{appendix}%
\usepackage{xcolor}%
\usepackage{textcomp}%
\usepackage{manyfoot}%
\usepackage{booktabs}%
\usepackage{algorithm}%
\usepackage{algorithmicx}%
\usepackage{algpseudocode}%
\usepackage{listings}%
\usepackage{cleveref}
\usepackage{a4wide}

\theoremstyle{thmstyleone}%
\theoremstyle{thmstyletwo}%

\theoremstyle{thmstylethree}%

\begin{document}

\title[Article Title]{A fully-implicit solving approach to an adaptive multi-scale model - coupling a vertical-equilibrium and full-dimensional model for compressible, multi-phase flow in porous media}


\author[1]{\fnm{Ivan} \sur{Buntic}}\email{ivan.buntic@iws.uni-stuttgart.de}
\author[1]{\fnm{Martin} \sur{Schneider}}\email{martin.schneider@iws.uni-stuttgart.de}
\author[1]{\fnm{Bernd} \sur{Flemisch}}\email{bernd.flemisch@iws.uni-stuttgart.de}
\author[1,2]{\fnm{Rainer} \sur{Helmig}}\email{rainer.helmig@iws.uni-stuttgart.de}

\affil[1]{\orgdiv{Department of Hydromechanics and Modelling of Hydrosystems}, \orgname{University of Stuttgart}}


\abstract{Vertical equilibrium models have proven to be well suited for simulating fluid flow in subsurface porous media such as saline aquifers with caprocks. However, in most cases the dimensionally reduced model lacks the accuracy to capture the dynamics of a system. While conventional full-dimensional models have the ability to represent dynamics, they come at the cost of high computational effort. We aim to combine the efficiency of the vertical equilibrium model and the accuracy of the full-dimensional model by coupling the two models adaptively in a unified framework and solving the emerging system of equations in a monolithic, fully-implicit approach. The model domains are coupled via mass-conserving fluxes while the model adaptivity is ruled by adaption criteria. Overall, the adaptive model shows an excellent behaviour both in terms of accuracy as well as efficiency, especially for elongated geometries of storage systems with large aspect ratios.}




\maketitle

\section{Introduction}
Green energy carriers play a crucial role in today's society as countries strive to move away from fossil fuels in order to secure a sustainable future by thwarting man-made climate change causes. Green gases are promised climate-friendly alternative energy carriers for covering the immense energy demands of countries' households and industry. As the energy requirements fluctuate during the course of a day, energy reservoirs are required to be able to match the change in energy consumption accordingly. To this end, there are mainly three options for large scale gas storage. While artificial salt caverns are already utilized for storing methane \cite{Haddenhorst1989} and hydrogen \cite{Caglayan2020}, their capacity is quite small in comparison to the other two options. Salt caverns offer roughly 7\% of the actively used global storage capacity, while aquifers and depleted gas fields represent 12\% and 75\% respectively \cite{Zhang2017}. However, these fractions only depict the currently used storage capacities. According to \cite{Luo2022}, the potential storage capacities of saline aquifers range from 5 to 100 times the potential storage capacities of depleted oil and gas fields analysed. These estimates were made based on reservoirs in China and the USA. As saline aquifers can potentially hold the largest amounts of gas, they will serve as the modelling basis for this work.

Storing gases in aquifers has already been successfully realized in the form of Carbon Capture and Storage (CCS) \cite{Michael2010}. In CCS, carbon dioxide is injected into aquifers in order to reduce the amount of free carbon dioxide in the atmosphere. The main difference between CCS and storage of green energy carriers is the cyclicity of the storage process. On the contrary to CCS where the carbon dioxide is stored permanently, the green energy carriers are cycled between injection and extraction on a daily basis in order to satisfy the fluctuating energy demands \cite{Miocic2023}.

As storage of gas in aquifers occurs in the porous medium of the underground, controlling the environment and observing the process is not trivial. Therefore, simulations of the storage process can offer valuable insights which allow understanding and planning the process in more detail \cite{Celia2015}. However, the large dimensions of aquifers often require significant computational efforts in order to discretize the large domain and predict the fluid flow. To alleviate the computational requirements for such simulations, different approaches have been developed, one of them being the vertical equilibrium (VE) method \cite{Gasda2009}\cite{Nordbotten2011}.  While the VE method indeed decreases the computational effort by reducing the effective spatial dimension of the computational domain by one, in many scenarios it unfortunately lacks the required accuracy to capture the physical processes correctly \cite{Becker2017}. To this end, adaptive models have been introduced which utilize the conventional full-dimensional equations where necessary and the VE equations where accuracy allows it~\cite{Becker2018}. There, the authors utilized an Implicit Pressure and Explicit Saturation (IMPES) solver. As the IMPES approach is subject to time step restrictions imposed by the Courant-Friedrichs-Lewy (CFL) condition \cite{Courant1967}, a single run of a simulation may require many individual time steps. While the IMPES solver performs excellently for purely advection-driven fluid flow, the method becomes more restrictive with increasing impact of non-linear effects such as capillary pressure \cite{Franc16} or buoyancy effects. In combination with the large number of time steps, this can result in an inefficient algorithm.

In this work, we lay out how to develop an adaptive scheme that couples full-dimensional and VE subdomains while using a monolithic, fully-implicit solving approach. We test the model's capabilities by simulating an injection process of methane into a saline aquifer which includes impermeable lenses and by comparing the results to a full-dimensional reference solution.

We start by presenting the mathematical models in \cref{section_mathModels}. Here, we present the full-dimensional model in \cref{section_FDmodel} and the VE model in \cref{section_VEmodel,sec:vedetails}. Following that, details of the discretization are discussed in \cref{section_discretization} while \cref{section_coupling} introduces the utilized coupling scheme with the definition of the coupling fluxes. In \cref{section_adaptivity}, we look into the adaption criteria for switching between the models in both directions. Then in our results, \cref{section_results}, an isotropic scenario as well as an impermeable lens scenario are investigated before focusing on potential improvements of the adaptive model. Finally, we summarize our findings in \cref{section_conclusion} and lend additional insight into the derivation of local VE functions in \cref{section_zp_mass_derivative}.

\section{Mathematical models} \label{section_mathModels}
In this section, we consider mathematical models for describing two-phase flow in porous media. First, the classical model is presented which we refer to as the full-dimensional (FD) model, consisting of the mass balances of each phase and Darcy equations which are valid within the entire domain. Then, we describe the second model, which is the vertical equilibrium (VE) model. This model's effective spatial dimension is reduced by one compared to the full-dimensional model, since the VE equations describe vertically averaged quantities. Finally, we introduce an adaptive scheme coupling the FD and VE model. From now on, capital letters describe quantities on the coarse VE scale, lower-case letters describe quantities of the FD model, while lower-case letters with a tilde represent quantities on the fine scale of the VE model.

All approaches compute the solution of multi-phase fluid flow in porous media on the REV scale, which assumes that the simulation domain consists of Representative Elementary Volumes (REVs). We presume isothermal conditions while the fluid system is described via an immiscible, compressible two-phase fluid system which consists of a wetting phase, in our case brine, and a non-wetting phase, in our case some green gas. Furthermore, for both models the main directions of the computational domain are congruent with Cartesian coordinates. 

\subsection{Full-dimensional model} \label{section_FDmodel}
The full-dimensional model is a set of equations which includes the mass balance equations as well as the Darcy equations for each phase. Let $\Omega \subset{\mathbb{R}^d} \text{ with } d \in \lbrace 1,2,3\rbrace$ denote the computational domain. The mass balance of a compressible fluid through a porous medium can be described via
\begin{equation} \label{eq_fullD_mass}
    \frac{\partial \phi \varrho_\alpha s_\alpha}{\partial t} + \nabla \cdot \left(\varrho_\alpha \boldsymbol{u}_\alpha \right) = \varrho_\alpha q_\alpha, \quad \alpha\in \{w,n\}.
\end{equation}
Here, $\phi$ represents the porosity of the porous medium, $s_\alpha$ stands for the saturation of phase $\alpha$, $\varrho_\alpha$ describes the phase density, $\boldsymbol{u}_\alpha$ is the Darcy-velocity of phase $\alpha$ and $q_\alpha$ describes sink and source terms while $w$ denotes the wetting phase and $n$ the non-wetting phase. Furthermore, the gas phase is treated as an ideal gas \cite{Poling2001}.
According to the Darcy equations, the momentum balance for each phase can be modelled by
\begin{equation} \label{eq_fullD_momentum}
    \boldsymbol{u}_\alpha = -\boldsymbol{k} \lambda_\alpha \left( \nabla p_\alpha - \varrho_\alpha g_z \nabla z\right), \quad \alpha\in \{w,n\},
\end{equation}
where $\boldsymbol{k}$ is the intrinsic permeability tensor of the porous medium, $\lambda_\alpha$ is the phase-specific mobility, $p_\alpha$ describes the phase pressure, $g_z$ is the gravitational acceleration in vertical direction and $z$ describes the vertical position.

To close the system of equations, additional closure relations for the capillary pressure and the saturations are given. Firstly, by including a capillary pressure in our system, the phase pressures are related via
\begin{equation} \label{eq_fullD_capPress}
    p_c = p_n - p_w,
\end{equation}
where $p_c$ is the capillary pressure and equals the difference of phase pressures. The capillary pressure is computed using an empirical relation, in this case the Brooks-Corey \cite{brooksCorey1964} capillary pressure-saturation relationship
\begin{equation} \label{eq_BrooksCorey_pc}
    p_c = p_e s_{w,e}^{-1/\lambda} = p_e \left( \frac{s_w - s_{w,r}}{1-s_{w,r}-s_{n,r}} \right)^{-1/\lambda},
\end{equation}
with $p_e$ being the entry pressure. The residual saturations of the respective phases are denoted as $s_{w,r}$ and $s_{n,r}$, which are modelling parameters while $s_{w,e}$ is called the effective wetting-phase saturation. Finally, $\lambda$ represents a modelling parameter which is a measure for the uniformity of the medium's pore sizes.

As the saturations describe fractions of the pore space occupied by the respective phase, for the second closing condition, the phase saturations should sum up to one which is the total available pore space,
\begin{equation} \label{eq_saturationClosing}
    s_w + s_n \overset{!}{=} 1.
\end{equation}

\subsection{Vertical equilibrium model} \label{section_VEmodel}
The VE model is a dimensionally reduced model, with its computational domain being $\Omega_{//}\in \mathbb{R}^{(d-1)}$, and is derived by integrating \cref{eq_fullD_mass} and \cref{eq_fullD_momentum} over the vertical height of the domain. A detailed derivation of the VE model equations for compressible fluids can be found in \cite{Guo2014}. By integrating the mass balance over the vertical height of the domain, we obtain
\begin{equation} \label{eq_VE_mass}
		\frac{\partial \varrho_\alpha \Phi S_\alpha}{\partial t} + \nabla_{//} \cdot \left(\varrho_\alpha \boldsymbol{U}_\alpha \right) = Q_\alpha,
\end{equation}
with the set of integrated quantities
\begin{align} \label{eq_integratedQuantities}
	\Phi &= \int_{z_B}^{z_T}{\phi \, \mathrm{d}z},      &      \boldsymbol{U}_\alpha &= \int_{z_B}^{z_T} {\boldsymbol{u}_\alpha \, \mathrm{d}z} , \notag \\
	S_\alpha &= \frac{1}{\Phi} \int_{z_B}^{z_T}{\phi s_\alpha \,\mathrm{d}z},     &      Q_\alpha &= \int_{z_B}^{z_T}{\varrho_\alpha q_\alpha \, \mathrm{d}z}.
\end{align}
Additionally, $z_B$ and $z_T$ describe the height of the bottom and top domain boundary, respectively. The subscript ${}_{//}$ indicates quantities and operators that live on the reduced space of the VE model which is the hyperplane perpendicular to the z direction.

Similarly, integrating the Darcy velocities over the vertical height leads to the depth-integrated formulation
\begin{equation} \label{eq_VE_momentum}
		\boldsymbol{U}_\alpha = - \boldsymbol{K} \Lambda_\alpha \left( \nabla_{//} P_\alpha - \varrho_\alpha g_z \nabla_{//} z_B \right)
\end{equation}
for the Darcy equations, where $P_\alpha=p_\alpha(z_B)$ describes the pressure at the bottom of the domain.
Here, the depth-integrated quantities are defined as
\begin{align} \label{eq_coarseMobilities}
	\boldsymbol{K} &= \int_{z_B}^{z_T}{\boldsymbol{k}_{//} \,\mathrm{d}z}       &      \Lambda_\alpha &= \boldsymbol{K}^{-1} \int_{z_B}^{z_T}{\boldsymbol{k}_{//} \tilde{\lambda}_\alpha \, \mathrm{d}z}.
\end{align}
As described in \cite{Guo2014}, the integrated versions of \cref{eq_VE_mass} and \cref{eq_VE_momentum} were derived with the assumption of negligible spatial derivatives of the density.




\subsection{Details of the VE model}
\label{sec:vedetails}
In the following, we will discuss the details of the VE model, ranging from its assumptions to the interplay between the coarse and fine scale of the model up to special considerations for computing model-specific quantities.

One assumption for the derivation of the VE model is the hydrostatic pressure distribution in vertical direction. Consequently, if the VE model is applied to domains where this assumption is violated, the results will deviate significantly from the physically expected behaviour. Furthermore, the assumption of hydrostatic pressure distribution implies that the model assumes the phases to segregate instantly. For larger density differences between the two fluids in a two-phase flow system, the segregation of the two phases will occur more quickly due to increased buoyancy forces. As a result, a phase equilibrium state will be reached in a short amount of time which facilitates the usage of a VE model.

It is important to understand that the VE model operates on two levels. The solutions of \cref{eq_VE_mass} and \cref{eq_VE_momentum} live on the first level $\Omega_{//}$ which we will refer to as the coarse scale from now on. Another appropriate name would also be the column scale, as \cref{eq_VE_mass} and \cref{eq_VE_momentum} operate on columns which contain vertically averaged quantities. On the contrary, we also define the so-called fine scale of the VE model. The fine scale only operates in the vertical direction within a column, effectively increasing the dimension by order one compared to the coarse scale.  Additionally, the fine scale is purely virtual as it does not directly participate in the solving scheme and serves the purpose of reconstructing quantities in vertical direction given the solution from the coarse scale. This approach helps to reduce the computational cost by solving on the lower dimension and visualizing on the larger dimension. 

\subsubsection{Reconstruction of fine-scale quantities}
In the following, we will go over one iteration of the VE model to describe the modelling of the quantity reconstruction on the fine scale which operates on $\Omega \in \mathbb{R}^d$. We start with an initial solution which is represented by our primary variables on the coarse scale, with the primary variables being the wetting-phase pressure and the non-wetting-phase saturation. Using the coarse-scale wetting-phase saturation, a key quantity can be computed, which we call the gas plume distance, $z_p$. The gas plume distance is the location of the segregation point between the two phases which are in vertical equilibrium. Above the segregation point, there is the gas plume while below the segregation point there is brine. The two phases are in vertical equilibrium once the vertical phase fluxes are negligible. Under the assumption of constant porosity, we can balance the overall mass of brine phase within each element of the VE coarse scale and the integral of the reconstructed brine-phase saturation over the vertical height, expecting both to match:
\begin{equation} \label{eq_gasPlumeDistInt}
    \int_{z_B}^{z_T}{\tilde{s}_w(z, z_p) \, \mathrm{d}z} = S_w (z_T - z_B).
\end{equation}
Here, the non-linear \cref{eq_gasPlumeDistInt} can be solved numerically to derive $z_p$.
The saturation $\tilde{s}_w(z, z_p)$ refers to the wetting-phase saturation on the reconstructed fine scale of the VE model and does not describe the saturation of the FD model. For a derivation of \cref{eq_gasPlumeDistInt}, see \cref{section_zp_mass_derivative} of the appendix.

\begin{figure}[h!]
  \begin{center}
    \includegraphics[width=1.0\textwidth]{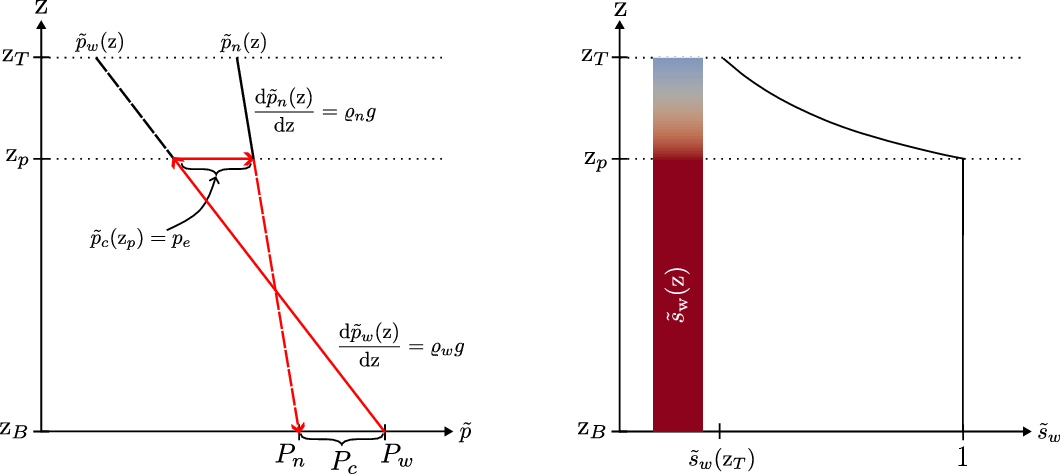}
  \end{center}
  \caption{\label{fig_reconstPlot} Graphical interpretation of reconstruction steps. (left) Sequence on deriving the coarse-scale capillary pressure is colored red and directions are marked with an arrow tip. Solid lines represent the profile of the reconstructed, fine-scale pressures, while dashed lines depict extrapolated values. The slopes of the reconstructed pressure lines depend on the respective phase density and gravitational constant. (right) Exemplary depiction of the reconstructed wetting-phase saturation within one VE column}
\end{figure}

In order to obtain an analytical formulation for the reconstructed saturation $\tilde{s}_w$, we first require relations for the reconstructed pressures and capillary pressure. Under consideration of a hydrostatic pressure profile, the fine-scale wetting-phase pressure $\tilde{p}_w(z)$ is computed as an extrapolation of the coarse-scale pressure, with the latter one being evaluated at the bottom $z_B$ of the domain. Given the coarse-scale density $\varrho_w = \varrho_w(P_w)$, we reconstruct the pressure via
\begin{equation} \label{eq_reconstPw}
    \tilde{p}_w(z) = P_w - \varrho_w g_z (z-z_B).
\end{equation}
For the gas-phase pressure reconstruction, we need to distinguish between the pressure within the gas plume $z > z_p$ which is affected by the gas density and the pressure below the gas plume $z\leq z_p$:
\begin{equation} \label{eq_reconstPn}
    \tilde{p}_n(z, z_p) =
    \begin{cases}
        P_w - \varrho_w g_z (z_p-z_B) + p_e - \varrho_n g_z (z-z_p),& z_p < z \leq z_T, \\
        P_w - \varrho_w g_z (z-z_B) + p_e,& z_B \leq z \leq z_p.
    \end{cases}
\end{equation}
The reconstructed capillary pressure can then be expressed as
\begin{equation} \label{eq_reconstPc}
    \tilde{p}_c(z, z_p) = \tilde{p}_n(z, z_p)-\tilde{p}_w(z) =
    \begin{cases}
        p_e + (\varrho_w - \varrho_n)g_z(z-z_p),& z_p < z \leq z_T, \\
        p_e,& z_B \leq z \leq z_p.
    \end{cases}
\end{equation}
By applying the inverse capillary pressure-saturation relation by Brooks-Corey, we obtain
\begin{align} \label{eq_reconstSw}
    \tilde{s}_w(z, z_p) &= p_c^{-1}(\tilde{p}_c(z,z_p)) \notag &\\
    &=
    \begin{cases}
        \left( p_e + \left( \varrho_w - \varrho_n \right)g_z\left( z-z_p \right)\right)^{-\lambda} p_e^{\lambda} \left( 1 - s_{w,r} - s_{n,r}\right) + s_{w,r},& z_p < z \leq z_T, \\
        1,& z_B \leq z \leq z_p,
    \end{cases}
\end{align}
as the analytical formulation for the reconstructed saturation.
Any other capillary pressure-saturation relation, e.g. by van Genuchten, is also valid.
Returning to the formulation in \cref{eq_gasPlumeDistInt} for the gas plume distance, the analytic form in \cref{eq_reconstSw} for the reconstructed saturation $\tilde{s}_w$ can be inserted and integrated, leading to
\begin{align*}
    \int_{z_B}^{z_T}{\tilde{s}_w(z) \, \mathrm{d}z} &= S_w (z_T - z_B) \\
    \int_{z_B}^{z_p}{\tilde{s}_w(z, z_p) \, \mathrm{d}z} + \int_{z_p}^{z_T}{\tilde{s}_w(z, z_p) \, \mathrm{d}z} - S_w (z_T - z_B) &= 0 \\
    z_p - z_B + \int_{z_p}^{z_T}{\tilde{s}_w(z, z_p) \, \mathrm{d}z} - S_w (z_T - z_B)  &= 0,
\end{align*}
and, finally,
\begin{align} \label{eq_gasPlumeDistIntExt}
    z_p - z_B &+ \frac{1}{1-\lambda} \left( p_e + \left( \varrho_w - \varrho_n \right)g_z\left( z_T-z_p \right)\right)^{1-\lambda} \frac{\left( 1 - s_{w,r} - s_{n,r}\right) p_e^{\lambda}}{\left( \varrho_w - \varrho_n\right)g_z} \nonumber \\
    &- \frac{p_e \left( 1 - s_{w,r} - s_{n,r} \right)}{\left( 1-\lambda\right) \left( \varrho_w - \varrho_n \right) g_z } + s_{w,r}\left( z_T - z_p \right) - S_w (z_T - z_B) & = 0.
\end{align}
Eq. \eqref{eq_gasPlumeDistIntExt} needs to be solved numerically for $z_p$ by using, for example, the Newton-Raphson method. In summary, \cref{fig_reconstPlot} gives an overview of the steps required to reconstruct the fine-scale pressure and to compute the coarse-scale capillary pressure, see the next section.

\subsubsection{Closing conditions}
Analogously to \cref{eq_fullD_capPress}, the coarse-scale capillary pressure 
\begin{equation} \label{eq_VE_capPress}
    P_c = P_n - P_w
\end{equation}
is defined as the difference between the coarse-scale pressures, where $P_n$ is the coarse-scale non-wetting-phase pressure, also evaluated at the bottom of the domain. On the contrary to the full-dimensional model, there is no need for an empirical relation like \cref{eq_BrooksCorey_pc} for the coarse-scale capillary pressure. While $P_w$ is known as one of the primary variables of the VE solution, we can utilize the assumption of hydrostatic pressure distribution to deduce the value of $P_n$ and thus simply compute $P_c$ as the difference, see \cref{fig_reconstPlot} for reference. At $z=z_p$, the reconstructed pressures differ exactly by the entry pressure $p_e$ implying that $\tilde{p}_c(z_p) = p_e$. Starting from $z_p$, we extrapolate $P_n$ by increasing the pressure value $\tilde{p}_n(z_p, P_w)=\tilde{p}_w(z_p, P_w)+\tilde{p}_c(z_p)$ linearly with the depth until we reach a value of $P_n=\tilde{p}_n(z_B, P_w) = \tilde{p}_w(z_p, P_w)+\tilde{p}_c(z_p) + \varrho_n g_z (z_p-z_B)$ at the bottom of the aquifer. Consequently, the coarse-scale capillary pressure in \cref{eq_VE_capPress} equals
\begin{align} \label{eq_PcCoarseValue}
    P_c = P_n - P_w &= \left(\tilde{p}_w(z_p)+p_e + \varrho_n g_z (z_p-z_B) \right) - \left( \tilde{p}_w(z_p) + \varrho_w g_z (z_p-z_B) \right) \nonumber\\
    &= \left( \varrho_n - \varrho_w \right) g_z \left( z_p-z_B \right) + p_e.
\end{align}
For large differences in densities, such as for brine and methane or brine and hydrogen, the coarse-scale capillary pressure $P_c$ usually has a negative value, which is difficult to interpret physically.
Simultaneously, as in \cref{eq_saturationClosing}, we obtain the constraint for the coarse-scale saturations 
\begin{equation}
    S_w + S_n \overset{!}{=} 1
\end{equation}
which directly follows from their definition in \cref{eq_integratedQuantities} and the constraint in \cref{eq_saturationClosing}.

\subsubsection{Evaluation of the remaining secondary variables}
It is worth mentioning that the coarse-scale densities $\varrho_w$ and $\varrho_n$ are functions of the pressure and temperature. We assume isothermal conditions and compute both coarse-scale densities by considering only the coarse-scale wetting-phase pressure. Usually, $\varrho_n$ should be computed by passing the coarse-scale non-wetting-phase pressure. However, as can be seen during the derivation of the coarse-scale quantities, there is a mutual dependency between $z_p$ and $\varrho_n$. For the computation of the gas plume distance $z_p$ in \cref{eq_gasPlumeDistIntExt}, $\varrho_n$ is required. Concurrently, the computation of $\varrho_n$ requires $P_n$ which in turn is dependent on $z_p$. One could either solve a system of non-linear equations to compute $\varrho_n$ and $z_p$ or simply compute $\varrho_n$ by using $P_w$ instead of $P_n$ as suggested by \cite{Nilsen2016}. Although this simplification leads to an error in $\varrho_n$, the deviation should not be significant given the large absolute values for the pressures in aquifers and also small relative differences in phase pressures compared to their absolute values.

For solving the VE equations, we still require the coarse-scale mobilities $\Lambda_\alpha$. To this end, the fine-scale mobilities $\tilde{\lambda}_\alpha$ are integrated over the height according to \cref{eq_coarseMobilities}. This is a local operation, meaning for one coarse-scale column only the fine-scale mobilities within that respective column are necessary to compute the coarse-scale mobility. By utilizing the definition
\begin{equation} \label{eq_effSaturation}
    \tilde{s}_{w,e} := \frac{\tilde{s}_w - s_{w,r}}{1-s_{w,r}-s_{n,r}}
\end{equation}
for the effective saturation $\tilde{s}_{w,e}$, we can formulate the empirical laws for the relative permeability-saturation relation
\begin{align} \label{eq_relPermSat}
  \tilde{k}_{rw} &= \left(\tilde{s}_{w,e} \right)^{3 + 2/\lambda} \\
  \tilde{k}_{rn} &= \left(1-\tilde{s}_{w,e}\right)^2 \left(1-\left( \tilde{s}_{w,e} \right)^{1 + 2/\lambda} \right)
\end{align}
by Brooks-Corey \cite{brooksCorey1964} and divide the relative permeabilities $\tilde{k}_{r\alpha}$ by their respective dynamic viscosity $\tilde{\mu}_\alpha$ to obtain the fine-scale mobilities
\begin{equation} \label{eq_mobility}
  \tilde{\lambda}_\alpha = \frac{\tilde{k}_{r\alpha}}{\tilde{\mu}_\alpha}.
\end{equation}
The dynamic viscosities $\tilde{\mu}_\alpha$ are computed via the coarse-scale pressure $P_w$ and according to the method of Chung \cite{Reid1987}. We calculate the representative mobility of a virtual, fine-scale element as the mean value of the mobility's integral from the fine-scale element's lower boundary to its upper boundary. Using the mean value, we obtain significantly better approximations of the coarse-scale mobilities compared to the values we obtained by evaluating the fine-scale mobility only at the element center. Having an accurate value for the coarse-scale mobility is crucial for a smooth and accurate solution.

\begin{figure}[htb]
  \begin{center}
    \includegraphics[width=0.75\textwidth]{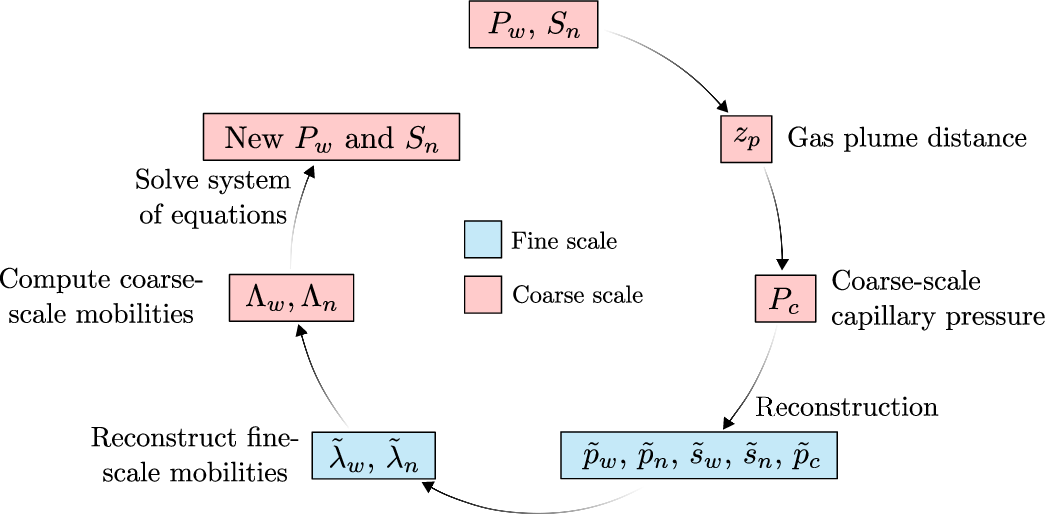}
  \end{center}
  \caption{\label{fig_flowchart} A flowchart depicting the general solving procedure within a VE scheme given the primary variables $P_w$ and $S_n$ on the coarse scale during initialization or at the beginning of a Newton solver iteration}
\end{figure}

Having all the necessary coarse-scale secondary variables, \cref{eq_VE_mass} and \cref{eq_VE_momentum} can be solved for a unique solution. 
In summary, \cref{fig_flowchart} shows the general solving procedure for the VE scheme. 
We start with given primary variables on the coarse-scale, representing, for example, the solution at a certain time step during the simulation or the deflected solution at an iteration step of the non-linear solver. In a next step, we compute the gas plume distance followed by the coarse-scale capillary pressure. Afterwards, all virtual, fine-scale quantities can be reconstructed via the coarse-scale primary variables and the gas plume distance. In the end, we obtain the coarse-scale mobilities by integrating the fine-scale mobilities according to \cref{eq_coarseMobilities}. At this point, all necessary coarse-scale secondary variables are known which are needed to solve the current system of equations. Figure~\ref{fig_flowchart} also clarifies, that the whole VE pipeline needs to be executed during each deflection step of the non-linear Newton-Raphson solver or in general whenever the coarse-scale primary variables are changed.

\subsection{Discretization details} \label{section_discretization}
All the presented models are discretized numerically via a
cell-centered finite volume method based on a structured grid, utilizing a two-point flux approximation. Other discretization schemes are discussed in \cite{Schneider2018}, while \cite{Schneider2017} presents alternatives for more complex grid layouts.

\begin{figure}[h!]
  \begin{center}
    \includegraphics[width=0.9\textwidth]{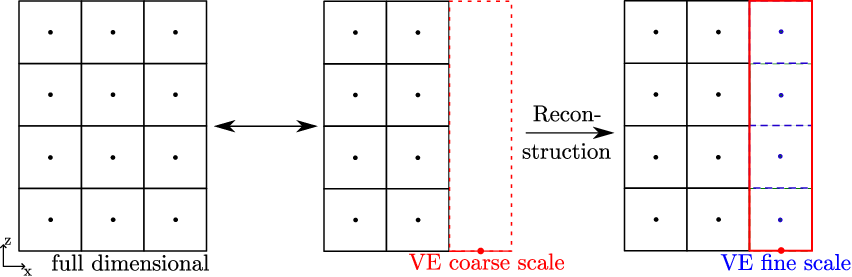}
  \end{center}
  \caption{\label{fig_discretizationComparison} This figure compares the discretization of the full-dimensional model and the VE model}
\end{figure}

As the fine-scale reconstruction of the VE model is continuous, a column of the VE model can be divided into an arbitrary number of fine-scale elements. In our case, we introduce a fine-scale discretization that matches the discretization of the FD model, see \cref{fig_discretizationComparison} for reference. There, on the left, a structured 2D grid for the full-dimensional model is shown. The middle grid represents the discretization after replacing one full-dimensional column with an element of the VE model. Strictly, this element only consists of the 1D interval at the bottom of the domain. The extrusion to a column by adding a height solely serves the purpose of visualization. From here on, we will use the words coarse scale columns and elements interchangeably. Nevertheless, the height is an essential parameter belonging to a coarse scale element. The right image shows a possible fine-scale discretization after reconstructing the coarse-scale VE solution from 1D to 2D. The reconstruction of the fine scale is a continuous operation, therefore the fine-scale discretization can be arbitrarily chosen. Here, we made a choice to match the discretization of the full-dimensional grid.

\subsection{Coupling scheme} \label{section_coupling}
Although the VE model leads to a significant speedup over the full-dimensional model, in many scenarios it generates wrong results. Therefore, we are interested in an adaptive scheme to balance the efficiency of the VE model and the accuracy of the full-dimensional model. As a first step, we have to discuss the details of the coupling between a FD and a VE subdomain before we can move on to the adaptive scheme. Similar to \cite{Becker2018}, we split our global, structured grid into subdomains of either the FD model or the VE model. Across the coupling interfaces of these subdomains, we presume a continuity of Darcy fluxes with the goal of conserving mass across coupling interfaces. Pursuing a fully implicit formulation, we cannot use total formulations of the coupling fluxes as proposed by \cite{Becker2018}, but need to resort to phase-specific coupling fluxes. The main challenge of the coupling scheme is the bridging of the two different-dimensional spaces the models are based on.

\begin{figure}[h!]
  \begin{center}
    \includegraphics[width=0.7\textwidth]{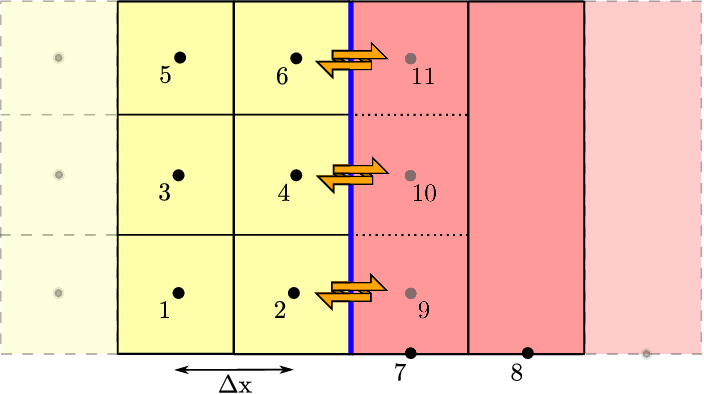}
  \end{center}
  \caption{\label{fig_couplingGraph} Drawing of the coupling scheme between a full-dimensional domain and a VE domain. The yellow subdomain represents the full-dimensional domain, while the red subdomain depicts the VE domain. Additionally, the blue line in the middle is the coupling interface which segregates the two subdomains. The flux from the left to the right domain is the sum of the individual fluxes, while the fluxes from the VE to the full-dimensional domain are considered individually. The horizontal discretization width is denoted as $\Delta \mathrm{x}$. The fluxes between coupled elements are illustrated as orange arrows}
\end{figure}

Illustrated in \cref{fig_couplingGraph}, the idea is to reconstruct fine-scale values within a coarse-scale column that mirror the discretization of the neighboring FD elements.
In \cref{fig_couplingGraph}, these would be the purely virtual degree of freedoms 9, 10 and 11. The flux from the full-dimensional subdomain to the VE domain is then computed as the sum of the individual fluxes from the coupled full-dimensional elements to their neighboring fine-scale VE elements. As depicted in \cref{fig_couplingGraph}, elements 2 and 9 are coupled, so are 4 and 10 as well as 6 and 11. Strictly speaking, the full-dimensional elements 2, 4 and 6 are actually coupled to the coarse-scale element 7. Introducing virtual fine-scale elements which mirror the discretization of the neighboring full-dimensional domain allows us to compute meaningful coupling fluxes.
Vice versa, the fluxes from the VE subdomain to the full-dimensional domain are computed individually from a coupled, fine-scale VE element to its neighboring full-dimensional neighbor.

In the following, we would like to lay out how an exemplary flux in 2D from element 2 to element 9 in \cref{fig_couplingGraph} is computed. For reasons of simplicity, the permeability tensor $\boldsymbol{k}$ is assumed to be isotropic thus reducing to a scalar $k$. Also the gradients $\nabla$ reduce to derivatives in x-direction $\frac{\partial}{\partial x}$ as we assume the full dimension to be 2D in this example. The coupling flux
\begin{equation}
    \boldsymbol{u}_{\alpha,c} = -k_c \lambda_{\alpha,c} \left( \nabla p_{\alpha,c} + \varrho_{\alpha,c}g\nabla\mathrm{z} \right)
\end{equation}
between elements 2 and 9 is computed by the means of a Darcy-flux computation as in eq.~(\ref{eq_fullD_momentum}) for both phases separately, where the density $\varrho_{\alpha,c}$ is an average of both neighboring densities $\varrho_{\alpha,c} = (\varrho_{\alpha,2}+\varrho_{\alpha,9})/2$, the pressure gradient $\nabla p_{\alpha,c}$ is the discretized derivative between the two neighboring pressures $\frac{p_9 - p_2}{\Delta \mathrm{x}}$, where $\Delta \mathrm{x}$ represents the horizontal discretization width.
Additionally, the permeability $k_c$ is computed as the harmonic average $k_c = \frac{k_9\cdot k_2}{k_9+k_2}$ of both neighboring permeabilities and the mobility is selected from the upwind direction. As the virtual, fine-scale element is always positioned at the same height as its full-dimensional neighbor, the gradient $\nabla\mathrm{z}$ is zero. As a result, the second entry of the flux vector $\boldsymbol{u}_{\alpha,c}$ is also zero. Therefore, in our Cartesian coordinate system the coupling flux is always perpendicular to the z-direction.

An additional remark is that one can debate how to correctly compute the density at the fine scale of the VE model. On the one hand, the pressure on the fine scale was reconstructed using the density of the respective coarse-scale column which is constant during one Newton iteration step. One could argue that this density should also be used on the fine scale in order to preserve mass between the fine and coarse scale. However, the coupling fluxes will be more accurate if fine-scale densities are used based on the reconstructed fine-scale pressures. As a result, the fine-scale densities will not be constant anymore within one column but will differ in vertical direction. In any case, the mass balance on the coarse scale is preserved, which is essential for the residual computation.

\subsection{Adaptivity} \label{section_adaptivity}
With an adaptive model, we strive to achieve a satisfying balance between the accuracy of the FD model and the efficiency of the VE model. To this end, we wish to utilize the full-dimensional model in parts of our simulation domain where accuracy demands it and the VE model otherwise. In other words, we deploy the VE model in parts where its reconstructed solution is close to the solution of the full-dimensional model.

For an adaptive model, additionally to the coupling scheme, we require adaption criteria. These rules are used to determine which model to use in which part of the domain. In our case, the adaption criteria are of relative nature. This means, we need to know what the current model of a domain is to deduce if a model switch is appropriate, global quantities alone as information are not sufficient.
The adaption criteria are divided into two groups: criteria for switching from a FD model to a VE model and vice versa. It is also worth to note, that these criteria are always computed within columns of a subdomain and at the beginning of each time step. In the full-dimensional context a column is represented by the vertical stack of elements that lie on top of each other, while in the VE context a column describes a planar element at the bottom of the domain which is extruded to the top of the domain along the z-axis.  

\subsubsection{Adaption criteria from the full-dimensional model to the VE model}
When analysing whether a full-dimensional column should be converted to the VE model, we compare the vertical saturation profile in this column with a virtual reconstructed saturation profile, as suggested by \cite{Becker2018}. If the difference between the two profiles is smaller than a threshold $\varepsilon_{crit}$, we can convert the column which would also mean that the error we introduce by replacing a full-dimensional column with a VE column is tolerable.

\begin{figure}[h!]
  \begin{center}
    \includegraphics[width=1.0\textwidth]{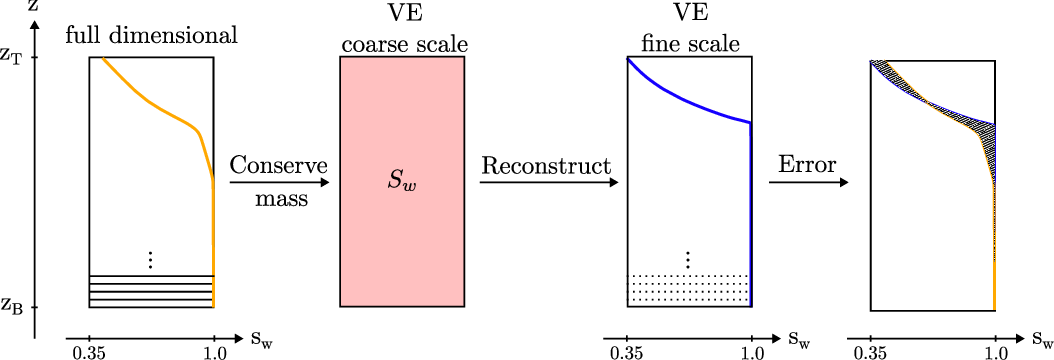}
  \end{center}
  \caption{\label{fig_criterionFromFullDToVE} Illustration of the saturation-based criterion computation. All depicted columns share the same spatial location and virtually lie on top of each other. For the FD column and the fine-scale VE column, the horizontal lines at the bottom indicate the discretization in vertical direction. Simultaneously, the coarse-scale VE column has no vertical refinement and only consists of one conservative value for the saturation. Finally, the error $\varepsilon_s$ in the far right graph is represented by the shaded area between the full-dimensional saturation profile and the virtual, fine-scale VE saturation profile}
\end{figure}

In a first step, we compute the non-wetting-phase mass $M_\text{FD}$ in a full-dimensional column by adding up the individual masses contained in the column's elements. In a second step, we determine the wetting-phase pressure at the bottom of this column, which acts as a virtual coarse-scale pressure $P_{w,v}$, where the subscript $v$ denotes a virtual quantity. Then, we can determine the mass conservative coarse-scale saturation $S_{n,v}$ by solving
\begin{equation} \label{eq_conservativeSFullDToVE}
     S_{n,v} \Phi \varrho_{n}(P_{w,v}) V_c = M_\text{FD},
\end{equation}
where $V_c$ describes the volume of the coarse-scale VE element. For a 1D coarse-scale element, the volume corresponds to the length of its element, which is $\Delta x$, in 2D the volume corresponds to the element's surface area. After solving for $S_{n,v}$, we have attained the two primary variables which describe the corresponding VE solution. From here, the virtual, fine-scale saturations $\tilde{s}_w$ are reconstructed at the heights of the full-dimensional elements which enables us to directly compare the values of $s_w$ and $\tilde{s}_w$ for each element of the full-dimensional column and to compute a column error according to
\begin{equation} \label{eq_critSatW}
  \varepsilon_s = \frac{1}{z_T - z_B} \int_{z_B}^{z_T}{\frac{\vert s_w - \tilde{s}_w \vert}{\vert s_w \vert} \, \mathrm{d}z}.
\end{equation}
We normalize by the absolute value of the FD saturation $\vert s_w \vert$ and the height of the column $z_T - z_B$. The normalization is especially useful, when computing criteria based on other quantities such as the pressure, to make the criteria more comparable. If $\varepsilon_s$ is smaller than a manually chosen threshold $\varepsilon_{crit}$, the column can be converted to the VE model without the introduction of a significant error in the solution. The concept of computing a criterion value $\varepsilon_s$ is visualized in \cref{fig_criterionFromFullDToVE}. 

Furthermore, we are interested in modelling the tip of the gas plume with the full-dimensional model as it is difficult to change the front back to FD once it is modelled via the VE model. As the tip of the plume is also able to fulfill the condition $\varepsilon_s < \varepsilon_{crit}$, this criterion alone is not sufficient. In addition to \cref{eq_critSatW} we also demand that $M_\text{FD}$ is larger than a manually chosen threshold $M_\text{crit}$. Finally, once $\varepsilon_s < \varepsilon_{crit}$ and $M_\text{FD}<M_\text{crit}$ are both fulfilled, a full-dimensional column will be converted to the VE model.

\subsubsection{Adaption criteria from the VE model to the full-dimensional model}
In the past, \cite{Becker2018} utilized proximity rules to decide when to convert a VE column back to the full-dimensional model. While this is an elegant approach for the IMPES scheme, as the CFL condition guarantees that the flow field will not travel further than one element during one time step, this constraint is not given in a fully-implicit environment. One can estimate the travel distance of the solution by taking into account the maximum Darcy velocity of the flow field and multiplying this maximum with the time step size, though this approach alone did not return consistent results. The true spread of the plume was either under- or overestimated by this approach.

There are two main criteria and one post-processing step for deciding whether a VE column will be converted to the full-dimensional model. The first criterion deals with the non-wetting phase mass $M_\text{VE}$ and checks whether the norm of the gradient of $M_\text{VE}$ in lateral direction, $\vert\vert \nabla_{//} M_\text{VE}\vert\vert_2$, is smaller than a threshold $M_\text{der,crit}$. The second constraint checks if the norm of the gradient of the coarse-scale wetting-phase saturation in lateral direction, $\vert\vert \nabla_{//} S_w\vert\vert_2$, is smaller than another threshold $S_\text{der,crit}$. In the results section, we will lay out the chosen values for all thresholds.  If any of the constraints $\vert\vert \nabla_{//} M_\text{VE}\vert\vert_2<M_\text{der,crit}$ or $\vert\vert \nabla_{//} S_w\vert\vert_2<S_\text{der,crit}$ is violated, a VE column should adopt the full-dimensional model.

As a post-processing step, we additionally estimate the travel distance of the gas plume by finding the maximum Darcy velocity in the domain, which should always be located at the injection area, and multiplying it by the time step size. Then, we extend the full-dimensional subdomain which discretizes the tip of the plume by the estimated travel distance, thus enlarging the right-most full-dimensional subdomain. To be on the safe side, we introduce a relaxation factor of 1.5, multiply the estimated travel distance by this factor and keep the elements in the new estimated travel distance full-dimensional. By doing so, we aim to consistently model the gas plume tip with the FD approach during time steps.

\section{Results} \label{section_results}
This section serves the purpose of obtaining first insights into the performance of our new, adaptive model and does not focus solely on the analysis of its efficiency. We intend to conduct a thorough efficiency analysis and comparison to an IMPES solving approach in a future work.
All numerical implementations and simulations for solving the model equations were implemented in DuMu\textsuperscript{x} \cite{koch21} \cite{dumux38}, which is a software-package for simulating flow in porous media utilizing a wide range of models and scales. The package is written in C++ and built as a module of the numerics environment DUNE \cite{dune21}.
The following simulations were run on a laptop with an AMD Ryzen 7 4700U CPU using eight cores with a maximum clock speed of 2000MHz. Furthermore, a memory of 14.8GiB was at disposal.
It is also worth mentioning that a code base developed for solving the full-dimensional equations can easily be adapted to solve the VE model \cite{Nilsen2016}. Specifically, only the computation of the porosities, sink/source terms, permeabilities, mobilities and boundary conditions needs to be modified to obtain a source code which solves the VE equations.

\begin{figure}[h!]
  \begin{center}
    \includegraphics[width=1.0\textwidth]{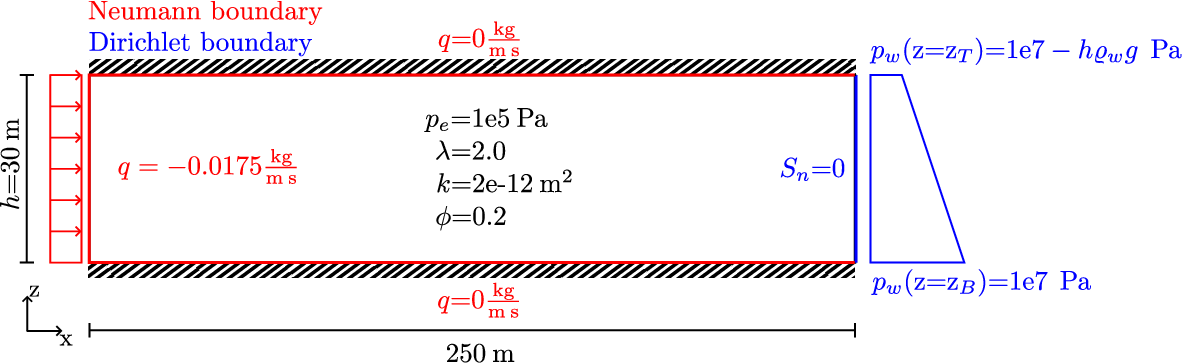}
  \end{center}
  \caption{\label{fig_isotropicScenario} Illustration of the investigated isotropic permeability scenario}
\end{figure}

We analyzed two scenarios to obtain a first insight into the performance of the adaptive model compared to the pure VE and pure FD model. To this end, an isotropic permeability scenario as well as an impermeable lens scenario are set up. As depicted in \cref{fig_isotropicScenario}, the modelled domain is 250m long and 30m high. On the top and bottom boundary, which are supposed to model cap rocks, we impose no-flow Neumann boundaries while we inject methane over the whole left boundary uniformly at a rate of $q=-0.0175 \frac{\text{kg}}{\text{m}\,\text{s}}$, a negative rate represents an injection process in our implementation. Finally, the right boundary is treated as a Dirichlet boundary with a hydrostatic pressure distribution. Initially, the whole domain is saturated with brine and we simulate for a duration of five days. Simultaneously, isothermal conditions were assumed with a temperature of $T=53^\circ\text{C}$. The values for the injection rate, temperature and other parameters are based on numerical experiments conducted in \cite{Becker2018}. Additionally, we model the underground reservoir as a rectangular 2D domain to obtain first insights into the performance of our new model. Extensions to 3D are currently in development. As we expect the injection area to be dominated by dynamic processes, we always assign the full-dimensional model there. By setting the Dirichlet conditions to the right boundary, we also expect the gas plume to not reach the right boundary during the simulation time, therefore the region around the right boundary is strictly modelled with the VE approach.

All values for the adaption criteria thresholds were chosen by trial and error. We found {$\varepsilon_{crit}=1.5\mathrm{e}\text{-}4 \, n_z$} to be a good estimate, where $n_z$ describes the number of vertical elements in a column. Furthermore, we chose {$M_\text{crit}=0.025\text{kg}$}, {$M_\text{der,crit}=3.0 \frac{\text{kg}}{\text{m}}$} as well as {$S_\text{der,crit}=0.0024 \frac{\text{1}}{\text{m}}$}. 

\subsection{Isotropic permeability scenario}
\begin{figure}[h!]
  \begin{center}
    \includegraphics[width=1.0\textwidth]{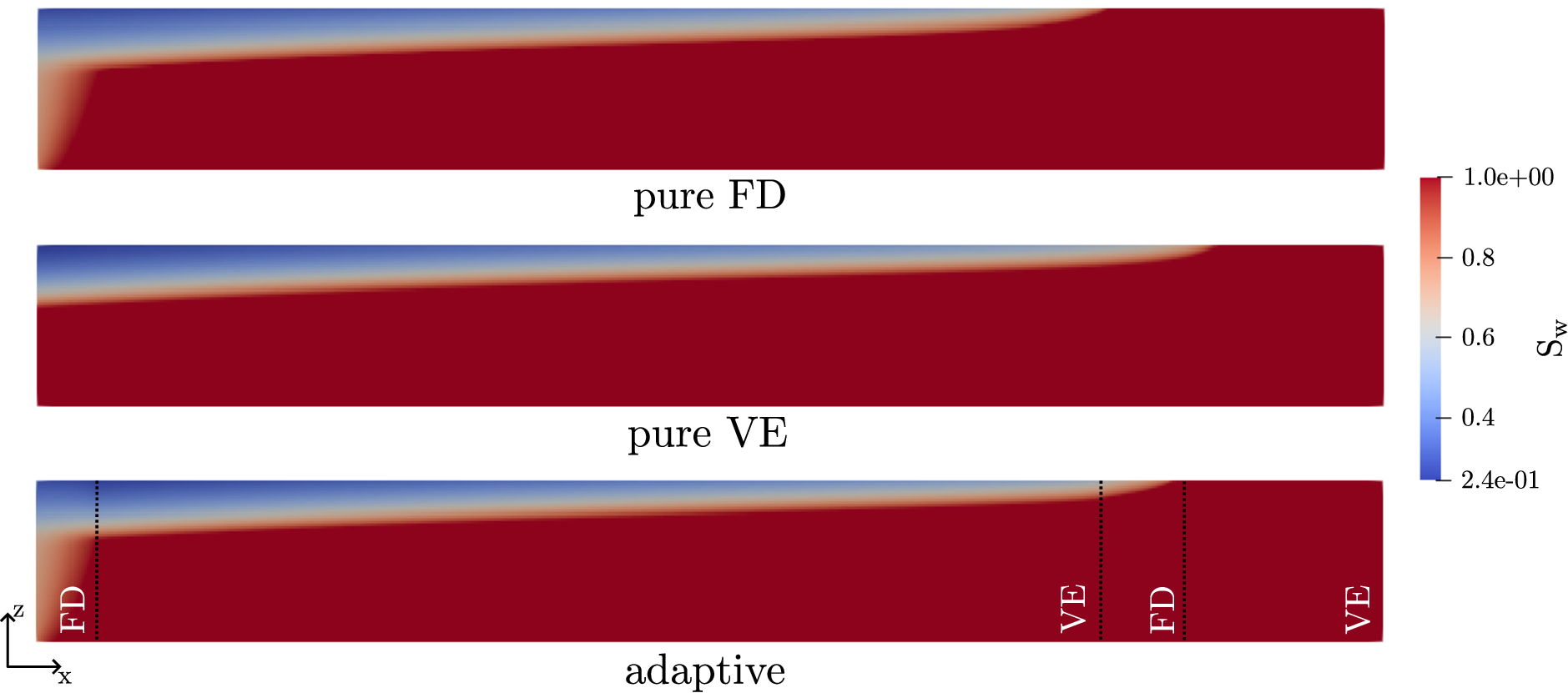}
  \end{center}
  \caption{\label{fig_isotropicResults} Isotropic test case: profiles of gas plumes after continuous injection of gas for five days. The full-dimensional and fine-scale VE model use a discretization of 500 elements in horizontal and 60 elements in vertical direction, while the coarse scale of the VE model uses a discretization of 500x1 elements. An impermeable lens is placed below the cap rock}
\end{figure}

For this scenario, the soil properties are distributed uniformly throughout the whole domain, see \cref{fig_isotropicScenario}. As the gas is injected continuously over the left boundary, we expect the gas to rise due to buoyancy forces. When it reaches the cap rock at the top, the gas accumulates there and then moves laterally in direction of the injection. We can observe this behaviour in the results of the full-dimensional solution in \cref{fig_isotropicResults}. Although the adaptive model reflects this behaviour well, the gas plume travelled further than in the full-dimensional solution. Meanwhile, the pure VE method fails to capture the buoyancy effects in well proximity. Also here, the extent of the gas plume is larger than in the full-dimensional model. However, at a first glance the main body of the gas plume is represented quite well. Table~\ref{table_isotropicResults} confirms our observations as the pure VE model exhibits the largest errors. The displayed errors are relative Root Mean Squared Errors (RMSE) of the wetting-phase saturation values while the reference solution was obtained from the full-dimensional results of a finely resolved discretization. When computing the element wise difference in the solution, we consider the boundaries of a coarse grid's element and inspect which elements of the reference grid are contained within these boundaries. Once we obtain these reference elements, we compute the average over them and consequently use this average to compute the error for the respective coarse grid element. Therefore, it is most convenient if the number of reference elements in each main direction is an integer multiple of the respective coarser grid's number of elements in that direction. In the end, we divide this absolute error value by the root mean integral over the reference saturation to obtain the relative error. For clarification, there are also error values displayed for the full-dimensional model, these errors are not equal to zero because of differing discretizations compared to the reference discretization.

By looking at the runtimes depicted in ~\cref{table_isotropicResults}, it is apparent that the full-dimensional model is the most expensive one for this scenario. While the adaptive method requires around half or even less of the full-dimensional model's computational time, the pure VE method only requires small fractions of the full-dimensional model's time. It is also worth mentioning, that for this specific scenario the adaptive model establishes three coupling interfaces, see \cref{fig_isotropicResults}. Later on, we will demonstrate that the efficiency and runtime of the adaptive model scales inversely to its number of coupling interfaces. This is also expected as the computational effort for the coupling scheme increases the more coupling interfaces are present.

Although the pure VE method is not able to resolve strongly dynamic processes, it can be a valid candidate to obtain quick results as well as an idea of what the process might look like. One could even argue, that the results of the pure VE model are good enough for the amount of saved computational time. However, this only holds true for simple, uniform scenarios as displayed in \cref{fig_isotropicResults}.
On the other hand, the adaptive model captures the dynamics of the process well by successfully distinguishing between the appropriate models at different parts of the domain. While the extent of the plume is still overestimated, a noticeable speedup over the full-dimensional model can also be observed.
When comparing the error values of the models, see the last column of \cref{table_isotropicResults}, for this simple scenario the adaptive model is more accurate than the pure VE model but not as accurate as the full-dimensional method with a matching grid resolution. The main error of the adaptive method is due to the poor prediction of the plume tip's location.

\begin{table}[h!]
\begin{tabular}{||c c | c c||} 
 \hline
 Model & Discretization elements & CPU time [s] & Error  \\ [0.5ex] 
 \hline\hline
 pure full-dimensional & 500x30 & 466.492 & 0.0089 \\ 
 \hline
 adaptive & 500x30 & 224.802 & 0.0300 \\
 \hline
 pure VE & 500x30 & 52.032 & 0.0436 \\
 \hline\hline
 pure full-dimensional & 500x60 & 1609.55 & 0.0096 \\ 
 \hline
 adaptive & 500x60 & 645.305 & 0.0306 \\
 \hline
 pure VE & 500x60 & 88.1739 & 0.0440 \\
 \hline
\end{tabular}
\caption{Runtime and error comparison between the different models for varying discretizations. The displayed error represents the relative RMSE of the wetting-phase saturations.}
\label{table_isotropicResults}
\end{table}

\subsection{Impermeable lens scenario}
\begin{figure}[h!]
  \begin{center}
    \includegraphics[width=1.0\textwidth]{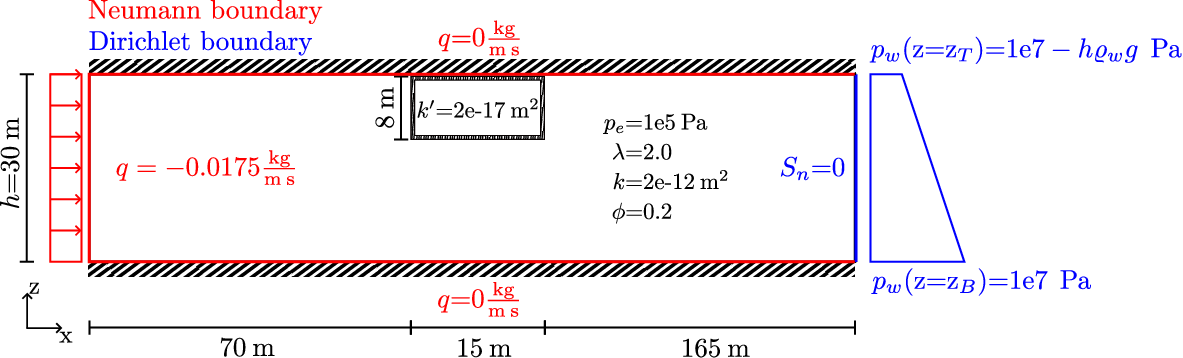}
  \end{center}
  \caption{\label{fig_lensScenario} Illustration of the investigated impermeable lens scenario}
\end{figure}

The initial and boundary conditions as well as the domain dimensions stay exactly the same as in the previous isotropic scenario. The sole difference is the introduction of an impermeable lens with the permeability $k'$ at the top of the domain, so the travelling gas plume is forced to flow around the lens, see \cref{fig_lensScenario}. Here, we would again expect the gas plume to accumulate below the cap rock, expand laterally  until it reaches the lens, accumulate in front of the lens until the gas plume is at least as thick as the height of the lens and then continue flowing laterally. 

\begin{figure}[h!]
  \begin{center}
    \includegraphics[width=1.0\textwidth]{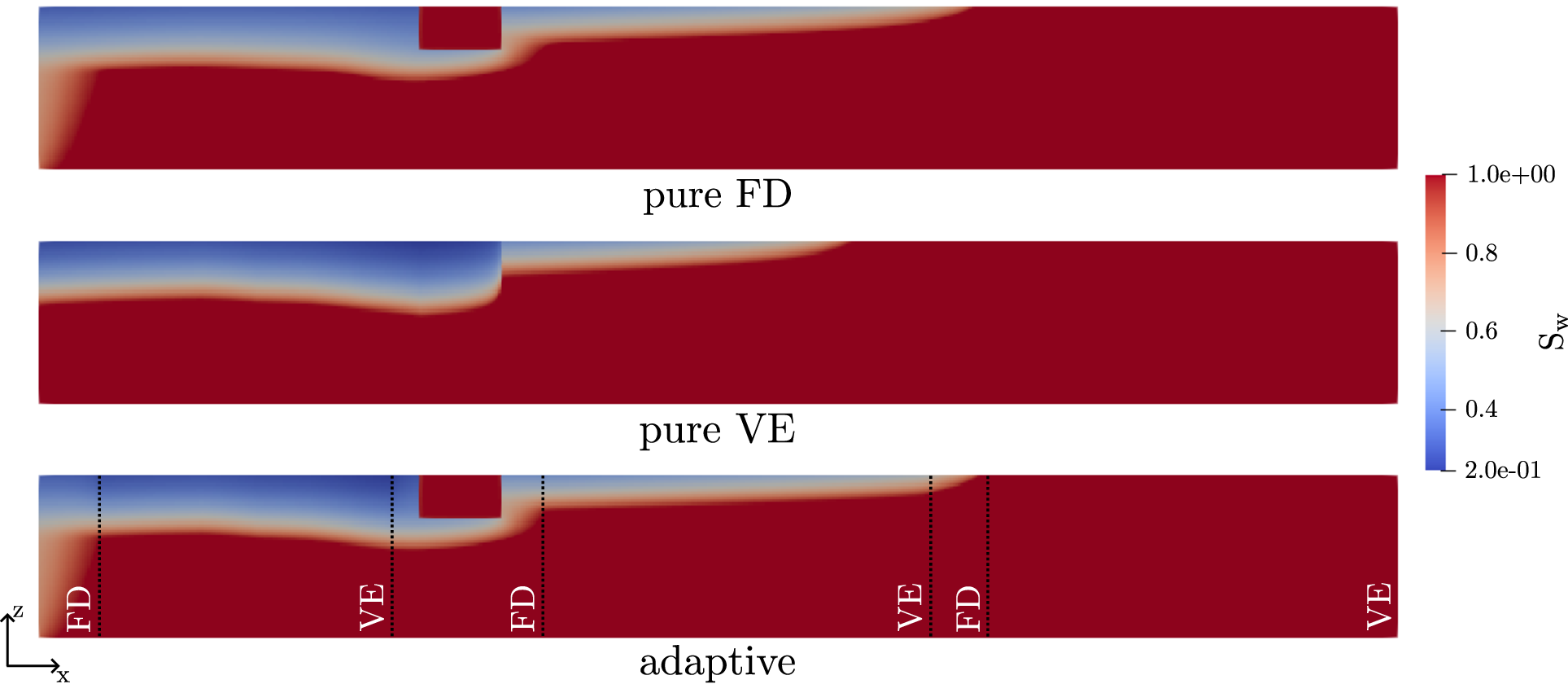}
  \end{center}
  \caption{\label{fig_impermeableResults} Simulation results after injecting gas for five continuous days over the left boundary while an impermeable lens is located at the top. For the fine scale a discretization of 500 horizontal and 60 vertical elements was chosen, while the coarse scale utilizes 500 horizontal cells and one vertical cell}
\end{figure}

The expected state after five days of injection can be seen in \cref{fig_impermeableResults} as depicted for the full-dimensional results. What stands out in this scenario is the inability of the pure VE model to capture the interactions around the impermeable lens. As the VE model averages its spatial quantities over the height, the impacts of the lens are neglected and local dynamics are not captured. It becomes apparent, that although the VE method is computationally very cheap, it lacks the ability to capture the vertical dynamics of a system. However, the VE method can be applied in conjunction with the full-dimensional model to obtain satisfying predictions, see the results of the adaptive model in \cref{fig_impermeableResults} in comparison to the full-dimensional ones. Even the extent of the gas plume tip is reflected well in this specific scenario which was not the case for the previous test case. The adaptive model is able to correctly apply the full-dimensional model at the injection area, at the lens and also the tip, thus all crucial parts to the dynamics can be modelled using the full-dimensional approach while the rest of the domain is modelled with the VE method.

However, by looking at \cref{table_lensResults}, we observe that the adaptive model is only faster than the full-dimensional model for finer discretizations. For certain discretizations, specifically those with a lower ratio of horizontal cells to vertical cells, the full-dimensional method tends to be faster than the adaptively coupled system. The reason for this is the expensive computational cost of the coupling scheme, as the scheme iterates over all coupled elements in vertical direction. During this process, quantities on the fine scale of the VE model are reconstructed before computing the coupling fluxes. With a finer discretization in vertical direction, more operations for the reconstruction are required. The additional step of reconstruction is necessary for each coupling interface which implies that the efficiency of the adaptive scheme does not only scale with the number of elements in vertical direction but also scales with the number of coupling interfaces. In general, the adaptive method shines in scenarios with a large ratio of the number of horizontal cells to vertical cells, meaning settings for which the computational effort is less dominated by the cost of the coupling scheme. The speedup over the FD model should be even more impactful for a 3D computational domain \cite{Nilsen2016}, as the ratio of full-dimensional elements to VE elements significantly decreases in a 3D setting. However, the extension to three dimensions was not the focus of this work and will be addressed in a future project.

\begin{figure}[h!]
  \begin{center}
    \includegraphics[width=1.0\textwidth]{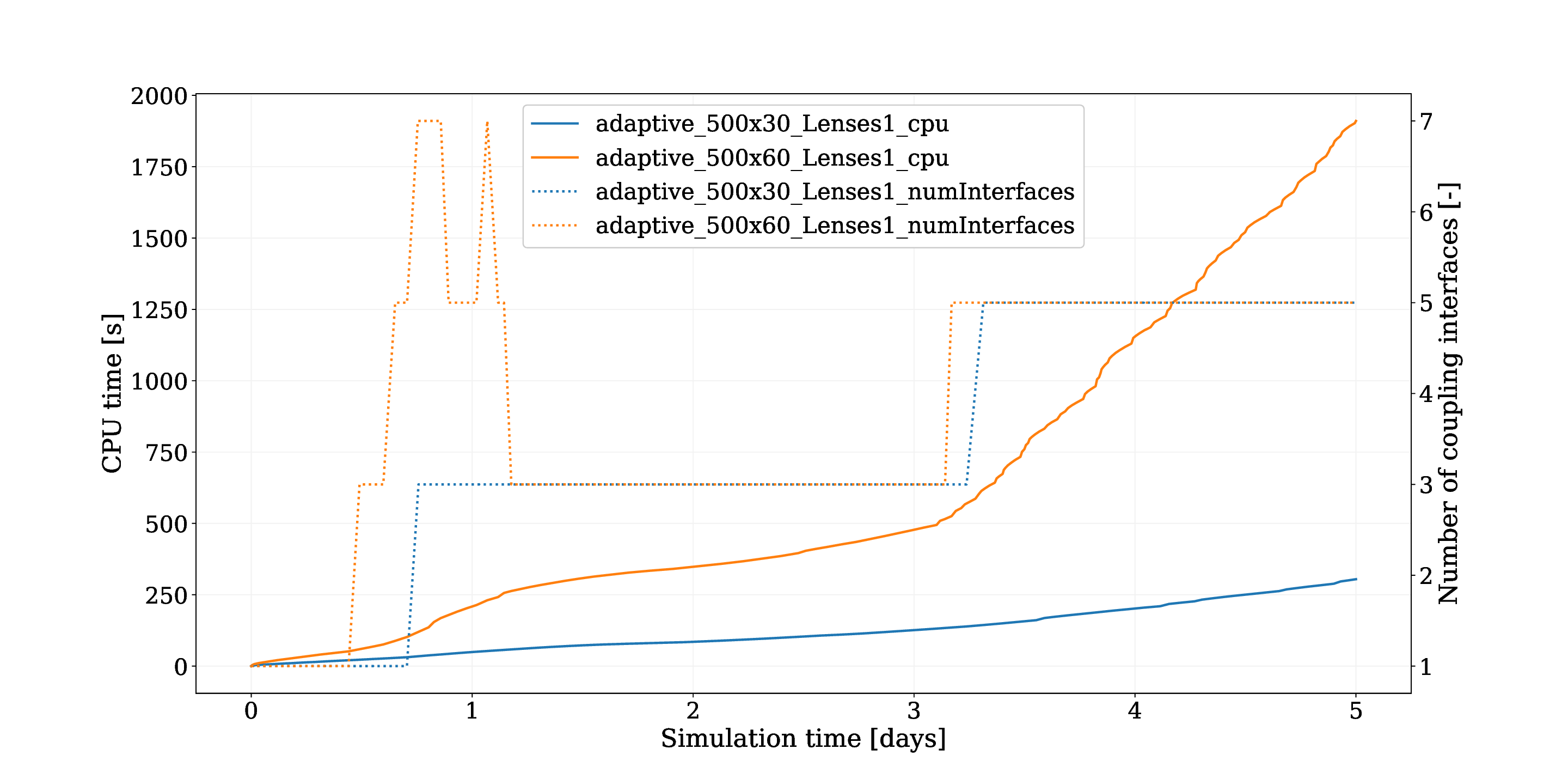}
  \end{center}
  \caption{\label{fig_numCouplingInterfaces} Lens scenario. The solid lines represent the CPU time over the simulated time while the dashed lines depict the number of coupling interfaces at a certain time}
\end{figure}

The impact of the presence of coupling interfaces on the adaptive model's efficiency is also illustrated in \cref{fig_numCouplingInterfaces}. The graph in \cref{fig_numCouplingInterfaces} can be divided into three regions. The first region stretches from the simulation start up to the point when three coupling interfaces are present, this is the period until the gas plume reaches the impermeable lens. The second region is depicted by the period for which three coupling interfaces are present. At this state, the domain around the impermeable lens is modelled with the full-dimensional approach. Finally, the third and last region is characterized by the presence of five interfaces, this occurs when the gas plume has overcome the impermeable lens and has travelled a considerable distance so another VE domain was formed between the tip of the plume and the lens. Two things can be deduced from this illustration. First, we observe that with an increase in the number of coupling interfaces the slope of the CPU time over the simulated time becomes steeper meaning the adaptive model requires more CPU time to simulate the same amount of time. This relation is understandable as with an increasing number of coupling interfaces, more coupling fluxes need to be computed. Unfortunately, computing the coupling fluxes is one of the most expensive operations in the adaptive model. The second observation are the oscillations in the number of coupling interfaces during the beginning of the simulation. In this specific scenario, the number of interfaces fluctuates between three and seven before stabilizing at three interfaces. These oscillations are an indicator that the adaptivity threshold $\varepsilon_{crit}$ is chosen poorly. Although, we previously introduced an a priori estimate of the threshold, it is an estimate after all and might require tweaking for individual scenarios. Nevertheless, the oscillations stabilize to three coupling interfaces. This is also the case for finer grids, where the range of oscillations is even larger but ultimately stabilizes to three coupling interfaces.

One can deduce that the ratio of the number of horizontal cells to vertical cells is a key measure for the efficiency of the adaptive model by comparing the runtimes for different discretizations in \cref{table_lensResults}. Ratios above 16 seem to favor the adaptive model in all of the inspected cases. However, depending on the total number of elements, ratios of around 8 can already be enough for the adaptive model to be faster than the full-dimensional approach while keeping the error low. Consequently, the ratio alone is not sufficient to predict which model choice should be favoured.

\begin{table}[h!]
\begin{tabular}{||c c c | c c||} 
 \hline
 Model & Discretization elements & Ratio cells & CPU time [s] & Error  \\ [0.5ex] 
 \hline\hline
 pure full-dimensional & 125x15 & 8.33 & 13.3494 & 0.0195 \\ 
 \hline
 adaptive & 125x15 & 8.33 & 13.8317 & 0.0266 \\
 \hline
 pure VE & 125x15 & 8.33 & 2.67414 & 0.0956 \\
 \hline\hline
 
 pure full-dimensional & 125x30 & 4.17 & 32.551 & 0.0212 \\ 
 \hline
 adaptive & 125x30 & 4.17 & 45.0887 & 0.0286 \\
 \hline
 pure VE & 125x30 & 4.17 & 3.64223 & 0.0959 \\
 \hline\hline

 pure full-dimensional & 250x15 & 16.67 & 40.7932 & 0.0117 \\ 
 \hline
 adaptive & 250x15 & 16.67 & 36.908 & 0.0213 \\
 \hline
 pure VE & 250x15 & 16.67 & 8.34679 & 0.1018 \\
 \hline\hline

 pure full-dimensional & 250x30 & 8.33 & 105.871 & 0.0137 \\ 
 \hline
 adaptive & 250x30 & 8.33 & 119.767 & 0.0209 \\
 \hline
 pure VE & 250x30 & 8.33 & 12.8261 & 0.1020\\
 \hline\hline

 pure full-dimensional & 250x60 & 4.17 & 312.679 & 0.0131\\ 
 \hline
 adaptive & 250x60 & 4.17 & 478.1 & 0.0207 \\
 \hline
 pure VE & 250x60 & 4.17 & 21.4736 & 0.1021 \\
 \hline\hline

 pure full-dimensional & 500x30 & 16.67 & 401.526 & 0.0082 \\ 
 \hline
 adaptive & 500x30 & 16.67 & 304.457 & 0.0187
 \\
 \hline
 pure VE & 500x30 & 16.67 & 47.5248 & 0.1042 \\
 \hline\hline

 pure full-dimensional & 500x60 & 8.33 & 2364.15 & 0.0055 \\ 
 \hline
 adaptive & 500x60 & 8.33 & 1910.24 & 0.0191 \\
 \hline
 pure VE & 500x60 & 8.33 & 81.1113 & 0.1043 \\
 \hline
\end{tabular}
\caption{Comparison of the CPU time and relative saturation error for different discretizations of the impermeable lens scenario. Additionally, the ratio of number of horizontal cells to the number of vertical cells in the domain is displayed for each discretization and labeled as "Ratio cells".}
\label{table_lensResults}
\end{table}

\begin{figure}[h!]
  \begin{center}
    \includegraphics[width=1.0\textwidth]{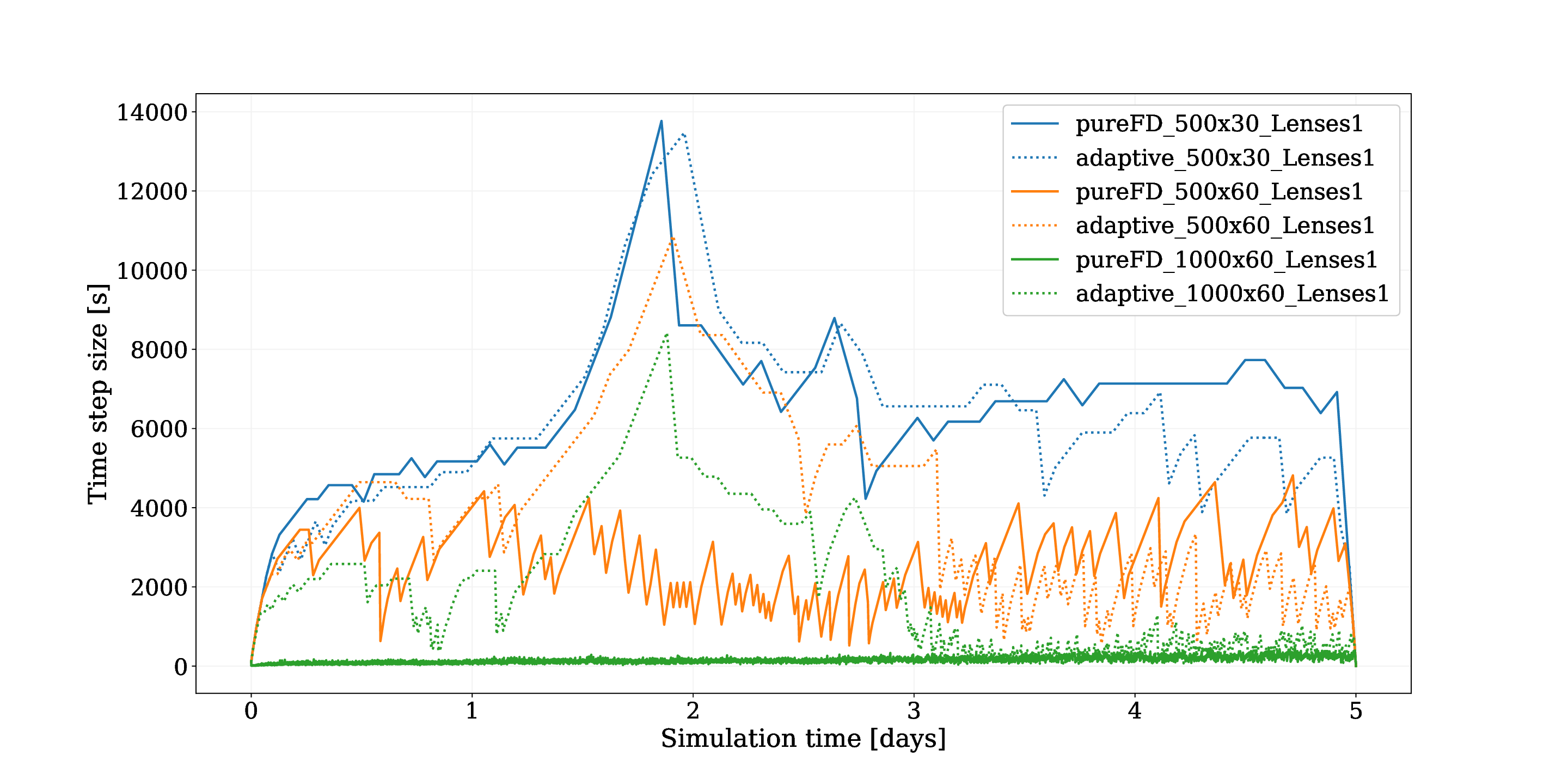}
  \end{center}
  \caption{\label{fig_lensStepWidths} Lens scenario. Comparison of the time step sizes for the adaptive and full-dimensional models plotted over the simulated time. The solid lines represent the full-dimensional model while the dashed lines belong the adaptive method}
\end{figure}

Figure~(\ref{fig_lensStepWidths}) lends additional insight into why the full-dimensional model takes longer than the adaptive model for a discretization of 500x60 even though the ratio of horizontal to vertical cells is only around eight. While for the discretization of 500x30 elements the time step sizes of both models show a peak at roughly two days, the time step sizes of the full-dimensional model for the finer discretization do not exhibit such a peak. Here, the time step sizes show a downward trend for increasing levels of refinement. On the other hand, the adaptive model manages to maintain the peak as well as the time step sizes leading to the peak even for the finer discretization which ultimately results in a faster simulation, regardless of the refinement in vertical direction. Refining the spatial dimension results in a finer resolution of gradients and non-linearities, these can become very steep which in turn can be challenging for a non-linear Newton solver. As the VE model computes its primary variables on the coarse mesh which introduces smearing effects, it is less subject to this issue and thus facilitates larger time step sizes also in the coupled, adaptive scheme.

Figure~(\ref{fig_lensStepWidths}) illustrates, how the time step sizes shrink with each spatial refinement for the full-dimensional model, while the adaptive approach is able to handle the time step size reduction better, as the VE model smears over fine-scale effects which can be challenging for the Newton solver otherwise. Nevertheless, for finer discretizations the adaptive model also tends towards smaller time step sizes in the later stages of the simulation. This is due to an increasing number of full-dimensional subdomains which at some point dominate the time step size choice. In our implementation, we do not restrict the time step size by some constraint but the Newton solver chooses it depending on the number of iterations needed for convergence. Due to the pure FD discretization resulting in a larger, more sparse matrix compared to the adaptive approach, it is likely that the pure FD matrix will have a worse condition number and therefore more iterations will be required during the iterative solving process of a Newton step. As a result of our solver's heuristic, the increased number of iterations will lead to generally smaller time step sizes for the pure FD discretization.

It is also worth mentioning that spatial mesh resolution alone does not necessarily lead to higher accuracy. As we do not keep our temporal discretization between the reference solution and the comparing solution consistent, it is possible for the temporal discretization error to become the dominant one, consequently a refinement in space does not guarantee a smaller error, as can be seen in \cref{table_lensResults}. To obtain a better comparison of the errors for the different spatial discretizations we would need to prohibit the non-linear solver to choose the time steps according to the mentioned heuristic but introduce a time step size constraint similar to the Courant-Friedrichs-Lewy number. However, this is not desired as a time step constraint would nullify the advantage of potentially large time steps of a fully-implicit solver. The effect of reduced time step sizes with finer spatial grids becomes even more apparent in case non-linearities such as capillary effects or in general diffusive effects dominate in a system. While for coarser grids, these effects are averaged over the coarser elements, for fine discretizations the non-linear impact becomes an increasing challenge for the non-linear solver.

It still remains to discuss, how the runtimes compare if the time step sizes were chosen similarly, would the FD model in that case be faster than the VE adaptive model. By having a look at the lines for the discretization of 500 times 30 cells in \cref{fig_lensStepWidths}, we can see that the FD and the adaptive model use roughly the same time step sizes at each point throughout the simulation. However, the respective entry in \cref{table_lensResults} shows that the adaptive model is faster overall by a considerable margin. Simultaneously, the entry for 125 times 30 cells in \cref{table_lensResults} illustrates that in this case the FD model is more efficient. Therefore, it is difficult to generalize which model is faster by solely looking at the time step sizes, the ratio of horizontal to vertical extent of the discretization but also the ratio of VE subdomains to FD subdomains plays an equally crucial role.

\subsection{Limits of the presented adaptive model}
\begin{figure}[h!]
  \begin{center}
    \includegraphics[width=1.0\textwidth]{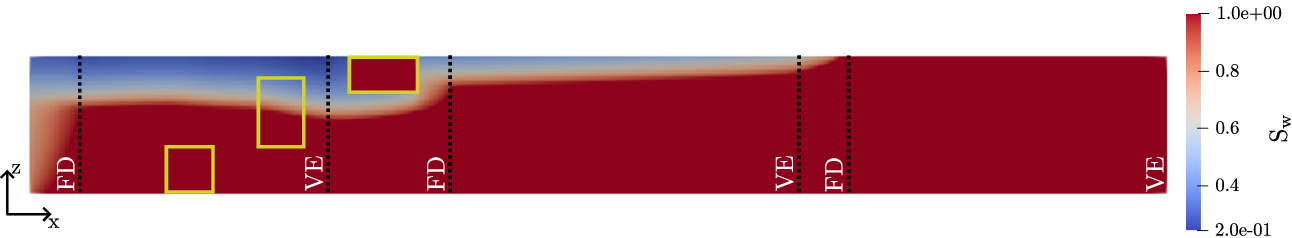}
  \end{center}
  \caption{\label{fig_threeLenses} Additional scenario with three impermeable lenses, one on the bottom of the domain, one in the middle and one on the top. The location of the lenses is marked with yellow boxes. The depicted solution belongs to the adaptive model}
\end{figure}

While the adaptive model shows promising results for maintaining the accuracy but also reducing the computational cost in the presented scenarios, the approach still has potential for improvement. In \cref{fig_threeLenses}, we introduce another scenario for which we have included three impermeable lenses throughout the domain highlighted via yellow boxes. The figure depicts the computed solution of the adaptive model under the same conditions as discussed in the previous test cases. However, by looking at the middle lens, we recognize that the presented solution does not match our expectations. We would expect the interior of the middle lens to remain saturated with brine. Instead, we observe that gas penetrates into the lens. The explanation is the following: as the gas rises and moves along the cap rock initially, at some point the plume tip reaches the top lens. At this stage the injection area and the tip are modelled via the full-dimensional approach while in-between a VE subdomain is formed. As the top lens hinders the gas plume from expanding, the gas accumulates in front of the top lens while gaining in thickness. During the thickening process, the gas plume reaches a certain thickness which already covers the middle lens partially. However, due the VE model being deployed in this area of the domain, the middle lens is not recognized as such but an average permeability over the height is used, thus ignoring the local effects of the middle lens. Our current model switching criteria do not catch this case. A possible option to avoid this issue would be to incorporate a modified version of the VE model for layered geological formations introduced in \cite{Guo2016} and \cite{Moyner2019} which could possibly allow the inclusion of lenses. However, this is not in the scope of this work anymore.

Another possible improvement to our adaptive model would be finding a more accurate prediction for the criterion threshold $\varepsilon_{crit}$, as the previously discussed oscillations in the number of coupling interfaces imply that the current value for the threshold might be slightly off.

While not the focus of this work, in the future we also intend to accelerate the reconstruction step of the VE model by introducing surrogate modelling or by utilizing multithreading. This should lead to a noticeable speedup as the reconstruction step for the coupling scheme constitutes the bottleneck of the adaptive scheme, which also explains the poor scaling of the adaptive scheme with the number of coupling interfaces.

\section{Conclusion} \label{section_conclusion}
We have introduced a coupled approach which adaptively deploys the full-dimensional model where accuracy demands it and the VE model where accuracy allows it. The emerging, monolithic system of equations is solved in a fully-implicit manner, thus allowing larger time step sizes compared to explicit solving approaches which are restricted by the CFL condition for stability purposes. To accommodate for the larger time step sizes of the implicit approach, we introduced additional adaption criteria for switching from a VE subdomain to a full-dimensional domain. Simultaneously, an empirical estimate for the value of the adaption criterion threshold has been proposed. While the time stepping scheme becomes increasingly more restrictive for the full-dimensional method when refining the discretizational grid in space, the adaptive approach seems to be more robust and maintains larger time step sizes also for finer grids. The reason for this is the smearing effect of the VE model's coarse scale, which averages dynamics over the vertical height thus ignoring local non-linearities which can be challenging for the non-linear solver.

We showed that the adaptive method's efficiency decreases with the number of coupling interfaces and also number of vertical elements but excels for larger ratios of the number of elements in horizontal direction compared to the vertical direction. In any case, the accuracy of the adaptive model exceeds the pure VE model's accuracy and comes close to the full-dimensional model's accuracy. Additionally, the adaptive model is capable of recognizing and capturing relevant dynamic effects, which are missed in the pure VE model.
Overall, for large-scale storage systems with significantly larger horizontal extension, the adaptive model retains the accuracy of the conventional, full-dimensional model while reducing the computational cost making it a valid candidate for computational studies of named storage systems.

\backmatter

\bmhead{Acknowledgements}

This work was funded by Deutsche Forschungsgemeinschaft (DFG, German Research Foundation) under Germany's Excellence Strategy - EXC 2075 – 390740016. We acknowledge the support by the Stuttgart Center for Simulation Science (SimTech).

\section*{Declarations}

\bmhead{Competing interests}
The authors have no competing interests to declare that are relevant to the content of this article.

\bmhead{Ethics approval and consent to participate}
Not applicable.

\bmhead{Consent for publication}
Not applicable.

\bmhead{Data availability}
Not applicable.

\bmhead{Materials availability}
Not applicable.

\bmhead{Code availability}
The code for reproducing the presented results can be accessed via \url{https://git.iws.uni-stuttgart.de/dumux-pub/buntic2024a}.

\bmhead{Author contribution}
Ivan Buntic: Methodology, Writing - original draft, Software; Martin Schneider: Methodology, Writing - review and editing, Software; Bernd Flemisch: Conceptualization, Methodology, Writing - review and editing, Funding acquisition; Rainer Helmig: Conceptualization, Methodology, Writing - review and editing, Funding acquisition. All authors have read and agreed to the published version of the manuscript.

\begin{appendices}

\section{Derivation of the mass balance for the gas plume distance}\label{section_zp_mass_derivative}

For deriving the mass balance relation for the computation of the gas plume distance, we assume that the mass within any vertical column $V$ is required to be conserved after the reconstruction step. Let $V_{//}$ denote the area of the column's cross section which is perpendicular to the vertical direction. For 1D, $V_{//}$ refers to the length of the interval, while for 2D it refers to the area of the cross section. Finally, the mass $M_\alpha$ of fluid phase $\alpha$ on the coarse scale is equivalent to
\begin{equation} \label{eq_massBalanceDer_coarse}
    M_\alpha = S_\alpha \Phi \varrho_\alpha V_{//}.
\end{equation}
For constant porosities, \cref{eq_massBalanceDer_coarse} simplifies to 
\begin{equation} \label{eq_massBalanceDer_coarse_simple}
    M_\alpha = S_\alpha  \phi  \left( z_T - z_B \right) \varrho_\alpha V_{//}.
\end{equation}
Balancing the coarse-scale mass and the fine-scale mass within the same column finally yields
\begin{equation} \label{eq_zp_mass_derivative}
    \int_{z_B}^{z_T}{\tilde{s}_\alpha \phi \varrho_\alpha V_{//} \, \mathrm{d}z} \overset{!}{=} S_\alpha \Phi \varrho_\alpha V_{//}.
\end{equation}
With the assumption of a constant porosity and the choice that the density within a fine-scale column is equal to the matching coarse-scale element, \cref{eq_zp_mass_derivative} simplifies to 
\begin{equation} \label{eq_zp_mass_derivative_simple}
    \int_{z_B}^{z_T}{\tilde{s}_\alpha \, \mathrm{d}z} \overset{!}{=} S_\alpha \left( z_T -z_B\right),
\end{equation}
which is equal to \cref{eq_gasPlumeDistInt} when inspecting the wetting phase.




\end{appendices}


\bibliography{sn-article}

\end{document}